%% file: main.tex
\documentclass{INTERSPEECH2023}

\interspeechcameraready 

\usepackage{bbm}
\usepackage{graphicx}
\usepackage{amsmath}
\usepackage{amssymb}
\usepackage{booktabs}
\usepackage{enumitem}
\usepackage{rotating}
\usepackage{colortbl}
\usepackage[table]{xcolor}
\usepackage{makecell}
\usepackage{tabu}
\usepackage{subcaption}
\usepackage{multicol}
\usepackage{multirow}
\usepackage{csquotes}
\usepackage{wrapfig}
\usepackage{url}

\title{Patch-Mix Contrastive Learning with Audio Spectrogram Transformer\\on Respiratory Sound Classification} 

\name{Sangmin Bae$^{1,5*}$\thanks{\hspace{-1.7em}*\,equal contribution \quad $^{\dagger}$corresponding authors \\\hspace*{0em}The code is available at: \url{https://github.com/raymin0223/patch-mix_contrastive_learning}}, June-Woo Kim$^{2,5*}$, Won-Yang Cho$^{3,5}$, Hyerim Baek$^{3,5}$, Soyoun Son$^{5}$,\\Byungjo Lee$^{4,5}$, Changwan Ha$^{5}$, Kyongpil Tae$^{5}$, Sungnyun Kim$^{1,5\dagger}$, Se-Young Yun$^{1\dagger}$}

\address{
  $^1$KAIST AI\quad
  $^2$Department of AI, Kyungpook National University\quad
  \\$^3$SmartSound\quad
  $^4$Dongguk University\quad
  $^5$MODULABS
  }
\email{\{bsmn0223, ksn4397, yunseyoung\}@kaist.ac.kr \quad kaen2891@knu.ac.kr}

\begin{document}

\maketitle

\input{tex/00_Abstract.tex}

\input{tex/01_Introduction.tex}

\input{tex/02_Related_Works.tex}

\input{tex/03_Preliminaries.tex}

\input{tex/04_Methodology.tex}

\input{tex/06_Results.tex}

\input{tex/07_Conclusion.tex}

\vspace{-3pt}
\section{Acknowledgements}
\vspace{-1pt}
This work was supported by the Ministry of Employment and Labor and HRD Korea's K-Digital Platform project, and Institute of Information \& communications Technology Planning \& Evaluation (IITP) grant funded by the Korea government (MSIT) (No. 2019-0-00075, Artificial Intelligence Graduate School Program (KAIST), 10\% and No.2022-0-00641, XVoice: Multi-Modal Voice Meta Learning, 45\%), and by the MSIT (Ministry of Science and ICT), Korea, under the ITRC (Information Technology Research Center) support program (IITP-2023-2020-0-01808) supervised by the IITP, 45\%.




\bibliographystyle{IEEEtran}
\bibliography{mybib}

\end{document}

%% file: tex/00_Abstract.tex
\begin{abstract}

Respiratory sound contains crucial information for the early diagnosis of fatal lung diseases. Since the COVID-19 pandemic, there has been a growing interest in contact-free medical care based on electronic stethoscopes. To this end, cutting-edge deep learning models have been developed to diagnose lung diseases; however, it is still challenging due to the scarcity of medical data. In this study, we demonstrate that the pretrained model on large-scale visual and audio datasets can be generalized to the respiratory sound classification task. In addition, we introduce a straightforward Patch-Mix augmentation, which randomly mixes patches between different samples, with Audio Spectrogram Transformer (AST). We further propose a novel and effective Patch-Mix Contrastive Learning to distinguish the mixed representations in the latent space. Our method achieves state-of-the-art performance on the ICBHI dataset, outperforming the prior leading score by an improvement of 4.08\%.
\end{abstract} 

\noindent\textbf{Index Terms}: Respiratory Sound Classification, ICBHI, Audio Spectrogram Transformer, Patch-Mix Contrastive Learning

%% file: tex/01_Introduction.tex
\section{Introduction}

Respiratory diseases, such as pneumonia or chronic obstructive pulmonary disease (COPD), impose a substantial burden on global health and well-being, constituting prevalent and common causes of death \cite{FIRS2021}. 
Furthermore, in the aftermath of the COVID-19 pandemic, there has been an increasing demand for accurate and expeditious diagnostic systems.
The conventional approach for detecting lung diseases is auscultation, which utilizes a stethoscope to listen to the respiratory sounds of patients in person \cite{arabi2021covid}. While hospitals can provide precise diagnoses, their effectiveness is significantly limited by the level of professional expertise and the subjectivity in interpretations\,\cite{sarkar2015auscultation, arts2020diagnostic}.

Recent developments in electronic stethoscopes provide recorded lung sounds, such as the ICBHI dataset \cite{rocha2018alpha}, increasing the interest in contact-free medical care and opening up new possibilities for automated lung sound examination.
Subsequently, numerous studies have developed methods for training neural networks that can achieve superior classification on the representative ICBHI dataset of stethoscopic sounds  \cite{ma2020lungrn+,yang2020adventitious, gairola2021respirenet, nguyen2022lung}.
Here, detection of abnormal respiratory sounds, such as \textit{crackle} and \textit{wheeze}, is critical for the accurate diagnosis of lung diseases \cite{bohadana2014fundamentals}.
In detail, crackles are characterized by discontinuous and non-tonal sounds that are associated with COPD, pneumonia, and lung fibrosis \cite{flietstra2011automated}.
Wheezes, on the other hand, are high-pitched continuous sounds that are commonly detected in patients with asthma and COPD \cite{bohadana2014fundamentals, reichert2008analysis}.


Despite its significance, detecting abnormal respiratory sounds suffers from the scarcity of medical data. 
The collection of adequate data to train deep networks still remains a challenge, and the usage of collected data is often restricted due to patient privacy concerns. Besides, the annotation of lung sounds requires specialized medical expertise.
Although previous literature has proposed various solutions, such as newly designed architectures \cite{ma2020lungrn+, nguyen2022lung, ma2019lungbrn}, data augmentation techniques \cite{ma2020lungrn+, gairola2021respirenet, wang2022domain}, or learning algorithms \cite{gairola2021respirenet, nguyen2022lung, moummad2022supervised}, these approaches have not yet achieved a satisfactory level of classification performance.
In this work, we present an effective learning approach that utilizes mixing augmentations on the pretrained Transformer model.

\begin{figure}[!t]
    \vspace{-2.5pt}
    \centering
    \includegraphics[width=0.87\linewidth]{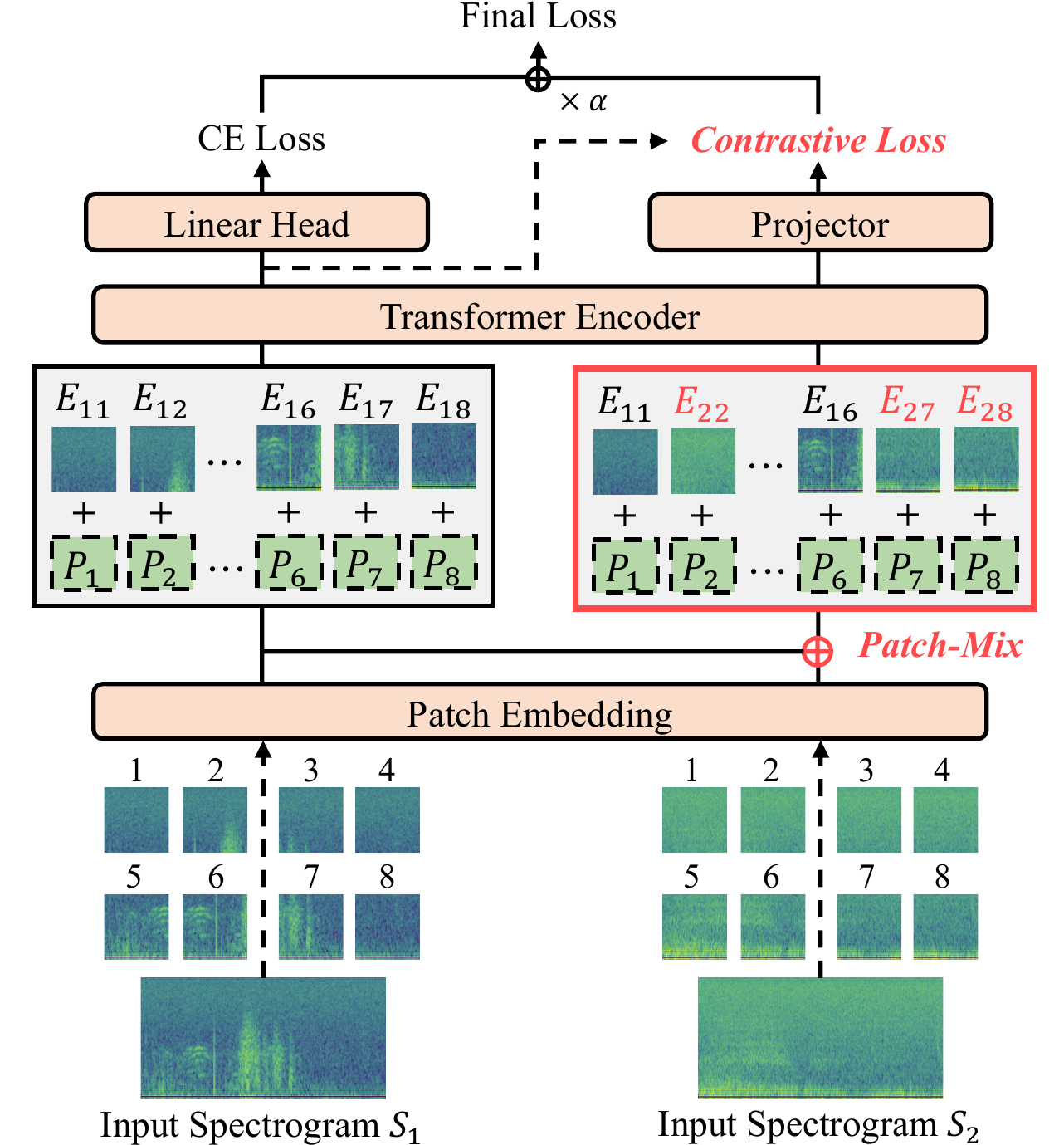}
    \vspace{-5pt}
    \caption{Overview of Patch-Mix Contrastive Learning. $E_{ij}$ denotes the embedding of $j$-th patch from the $i$-th spectrogram, and $P_j$ is $j$-th positional embedding. The \texttt{class} and \texttt{distill} tokens in the AST are omitted. Note that dashed lines indicate the stop-gradient operation.}
    \label{fig:overview}
    \vspace{-12pt}
\end{figure}

So far, prior research on lung sound classification has proved that fine-tuning the pretrained conventional visual models is a straightforward and effective approach to address the shortage of data \cite{gairola2021respirenet, nguyen2022lung, wang2022domain}.
However, there is a lack of study on recent attention-based models such as Audio Spectrogram Transformer\,(AST) \cite{gong2021ast}, which are being widely used in the audio domain \cite{gong2022ssast,baade2022mae}.
Moreover, most works have considered either only visual- or audio-pretrained models by ImageNet \cite{deng2009imagenet} and AudioSet \cite{audioset}, respectively.
Since we convert lung sounds to Mel-spectrograms for the image-level classification, the image and audio pretrained model can be generalized to the respiratory sound classification task.
Indeed, we demonstrate that fine-tuning an AST model, pretrained on large-scale datasets of both domains, achieves state-of-the-art performance in respiratory sound classification, especially on the ICBHI dataset.

In order to alleviate overfitting on insufficient samples of the ICBHI dataset, several studies have also introduced data augmentation techniques such as label-aware concatenation \cite{gairola2021respirenet} or Mixup \cite{wang2022domain}.
Taking advantage of the AST model that patchifies the input image, we propose a straightforward Patch-Mix augmentation that randomly mixes patches (or tokens) between different instances (see Figure\,\ref{fig:overview}).
However, respiratory sound classes contain a hierarchy such that when instances from the \textit{normal} and \textit{crackle} classes are evenly mixed, it is ambiguous to regard the mixture as equally distributed classes or just an abnormal crackle class. 
Therefore, the direct application of conventional Mixup \cite{zhang2017mixup} or CutMix \cite{yun2019cutmix} loss functions has limitations in the context of lung sound classification tasks.

To this end, we propose a novel Patch-Mix Contrastive Learning to distinguish mixed representations in the latent space, which enables the model to learn the extent to which the mixed representation resembles the original, irrespective of the label information \cite{kim2020mixco, lee2020mix}. It is clearly different from the conventional contrastive loss \cite{chen2020simple, he2020momentum} in that we consider representations from mixed sources as positive pairs.
In practice, our proposed methods achieved a new record 62.37\% Score on the ICBHI dataset, improving by 4.08\% compared to the previous best Score.
We believe that our methods can be extended to various benchmarks and downstream tasks.

%% file: tex/02_Related_Works.tex
\section{Related Works}

\subsection{Respiratory Sound Classification}
\vspace{-2pt}

Since the ICBHI 2017 challenge, numerous neural network-based methods have emerged for the lung sound classification task, particularly those that utilize or modify conventional residual block-based models \cite{ma2020lungrn+, yang2020adventitious, gairola2021respirenet,  nguyen2022lung, ma2019lungbrn, wang2022domain, li2021lungattn}.
Several studies have yielded promising results by utilizing pretrained models on ImageNet \cite{gairola2021respirenet, nguyen2022lung, wang2022domain} or AudioSet \cite{moummad2022supervised}, rather than training from scratch. 
Furthermore, to effectively address the issue of limited data, various learning protocols have been proposed, such as device-specific fine-tuning \cite{gairola2021respirenet}, task-specific co-tuning \cite{nguyen2022lung}, or supervised contrastive learning \cite{moummad2022supervised}. 
Our study first leverages the innovative AST model, which deviates from conventional visual models, and attains state-of-the-art performance on the ICBHI lung sound dataset. Additionally, our proposed contrastive learning framework has shown synergy with the Transformer model that patchifies the audio spectrograms.
\vspace{-5pt}

\subsection{Mixed Representation Learning}
\vspace{-2pt}

Due to a shortage of lung sound data, previous research has proposed augmentation techniques such as label-aware concatenation \cite{gairola2021respirenet} or Mixup in the respiratory audio \cite{ma2020lungrn+, wang2022domain}. Mixing different samples, such as Mixup \cite{zhang2017mixup} or CutMix \cite{yun2019cutmix}, is a popular augmentation technique that has been applied regardless of the modality. Recently, token-level CutMix methods combined with self-attention visual models have been proposed: TokenMix \cite{liu2022tokenmix} assigns the target score based on the content-based activation maps of two images, and TokenMixup \cite{choi2022tokenmixup} proposes Mixup with guided attention on the Transformer architectures. 
In this study, we introduce a straightforward Patch-Mix technique that replaces randomly selected patches with those from other samples without any complex operation. 
Drawing inspiration from prior works \cite{kim2020mixco, lee2020mix}, we further formulate a mixed contrastive loss to distinguish the mixture in latent space, learning how much the mixed representation resembles the original.

%% file: tex/03_Preliminaries.tex
\vspace{-5pt}
\section{Preliminaries} 

\subsection{Dataset Description}
\vspace{-2pt}

We used the respiratory sound database of ICBHI 2017 challenge organized at Int. Conf. on Biomedical Health Informatics \cite{rocha2018alpha}. The ICBHI dataset comprises 6,898 respiratory cycles, which have a duration of approximately 5.5 hours and are officially split into a train set (60\%) and a test set (40\%). Note that the patients are divided into either the train or test set without overlap. Each breathing cycle is annotated as one of the four classes: \textit{normal}, \textit{crackle}, \textit{wheeze}, and \textit{both} (crackle and wheeze).
The train set includes a total of 539 recordings taken from 79 patients, that consist of 1,215 cycles of crackles, 501 cycles of wheezes, 363 cycles of both crackles and wheezes, and 2,063 cycles of normal breathing.
Similarly, the test set includes 381 recordings taken from 49 patients, with a total of 649 cycles of crackles, 385 cycles of wheezes, 143 cycles of both crackles and wheezes, and 1,579 cycles of normal breathing. 
\vspace{-5pt}

\subsection{Preprocessing Details}
\vspace{-2pt}

Since all respiratory cycles in the ICBHI dataset have various sampling rates, we resampled all recordings to 16kHz.
Besides, to ensure the same length of all samples, we set each duration to 8 seconds, corresponding to 798 frames.
We truncated the samples for longer cycles while we duplicated the cycle with a fade in/out operation until the desired length \cite{moummad2022supervised}. 
Then, we converted the audio waveform to a sequence of 128-dimensional log Mel filterbank (Fbank) features, using the window and overlap size of 25ms and 10ms, respectively \cite{gong2021ast}.
We also applied the standard normalization on the spectrograms with the mean and standard deviation of --4.27 and 4.57, respectively.
\vspace{-5pt}

\subsection{Training Details}
\vspace{-2pt}

In order to prevent overfitting, we utilized SpecAugment \cite{park2019specaugment} with a maximum mask length of 160 frames for the time domain and 48 bins for the frequency domain. Here, we used the average values of each spectrogram for masking, and time-warping augmentation is excluded for simplicity.
For the AST model, we used the Adam optimizer with a learning rate of 5e--5, cosine scheduling, and a batch size of 8. We fine-tuned the pretrained AST model for only 50 epochs but trained for 200 epochs when starting from scratch.
For other architectures, we used a learning rate of 1e--3, a batch size of 128, and trained the model for 200 epochs. For the stable training, we applied the momentum update with the coefficient of 0.5 to all learnable parameters. Besides, except for when using mixing augmentation, we used a weighted cross-entropy loss, inversely proportional to the number of class samples \cite{gairola2021respirenet, moummad2022supervised}.
Note that we present the results of our experiments over five random runs.
\vspace{-5pt}

\subsection{Evaluation Metrics}
\vspace{-2pt}

Following the evaluation metrics of the ICBHI 2017 challenge, we evaluate the classification performance by Score, the average of specificity\,($S_p$) and sensitivity\,($S_e$).
The specificity and sensitivity are defined as follows: 
\vspace{-2pt}
\begin{equation}
S_p = \frac{C_n}{N_n}, \quad
S_e = \frac{C_c + C_w + C_b}{N_c + N_w + N_b}
\vspace{-2pt}
\end{equation}
where $C_i$ and $N_i$ are the number of correctly classified and total samples in class $i \in \{ normal, crackle, wheeze, both \}$, respectively.
In addition to the evaluation of the 4-class case, we also consider the 2-class scenario where crackle, wheeze, and both classes are grouped into a single class, labeled as \textit{abnormal}. For simplicity, we utilize a trained classifier for the 4-class and subsequently compute the $S_e$ for the 2-class version \cite{nguyen2022lung}.

%% file: tex/04_Methodology.tex
\section{Methodology}

\input{tex/04_1_Audio_Spectrogram_Pretraining.tex}
\input{tex/04_2_Patch-Mix_Contrastive_Learning.tex}

%% file: tex/04_1_Audio_Spectrogram_Pretraining.tex
\subsection{Audio Spectrogram Transformer}
\vspace{-2pt}

The respiratory sound classification task is a challenging problem due to the limited number of samples available to train the large networks. As a result, prior works have utilized pretrained conventional visual models on either ImageNet\,(IN) or AudioSet\,(AS) to overcome the scarcity of training samples. 
AudioSet contains over 2 million audio clips representing various sound types, even including respiratory sounds like breathing, cough, sneeze, or sniff \cite{audioset}. In fact, Moummad \textit{et al.}\,\cite{moummad2022supervised} achieved comparable results to the best Score using the AudioSet pretrained CNN6 model.

Accordingly, pretraining on both visual and audio domains would be a powerful technique for improving the performance of respiratory sounds. The Audio Spectrogram Transformer (AST) \cite{gong2021ast}, a purely attention-based model, is sequentially pretrained on ImageNet and AudioSet, demonstrating significant improvements in various audio downstream tasks. The self-attention mechanism also enables it to better capture long-range dependency in both time and frequency domains. In contrast to previous works that solely rely on convolution models, we adopted the \textit{base} AST model pretrained on both ImageNet and AudioSet for the lung sound classification task.

Table \,\ref{tab:arch_pretrain_comparison} summarizes the classification Score for the 4-class evaluation across different architectures and datasets. In order to confirm the influence of the architecture and pretraining datasets, we fine-tuned the models without additional learning techniques. Note that we compared only the reported values in previous studies that used the official split.
Although training the AST model from scratch is challenging, there was a significant improvement as it was more pretrained on large datasets, resulting in state-of-the-art performance with ImageNet and AudioSet pretraining. This suggests that pretraining on datasets from both domains indeed enhances the generalization of the AST for the lung sound classification task.
\vspace{-5pt}

\input{tables/arch_pretrain_comparison.tex}

%% file: tables/arch_pretrain_comparison.tex
\begin{table}[!t]
    \centering
    \caption{Lung sound classification performance according to various architectures and pretraining datasets. All reported numbers are the results of simple fine-tuning on the ICBHI dataset. IN and AS indicate ImageNet \cite{deng2009imagenet} and AudioSet \cite{audioset}, respectively. $\dagger$ denotes the experimental results of our implementation. \textbf{Best} and {\underline{second best}} results.
    }\label{tab:arch_pretrain_comparison}
    \vspace{-2pt}
    \addtolength{\tabcolsep}{0pt}
    \resizebox{\linewidth}{!}{
    \begin{tabular}{lclll}
    \toprule
    architecture & pretrain & $S_p$\,(\%) & $S_e$\,(\%) & \textbf{Score}\,(\%) \\
    \hline \midrule
    %
    $\text{EfficientNet}^\dagger$ & - & $\text{70.92}_{\pm 3.61}$ & $\text{28.93}_{\pm 3.05}$ & $\text{49.93}_{\pm 0.59}$ \\
    $\text{ResNet18}^\dagger$ & - & $\text{70.68}_{\pm 3.28}$ & $\text{30.14}_{\pm 3.98}$ & $\text{50.41}_{\pm 0.60}$ \\
    $\text{bi-ResNet}$ \cite{ma2019lungbrn} & - & 69.20 & 31.12 & 50.16 \\
    $\text{ResNet-NL}$ \cite{ma2020lungrn+} & - & 63.20 & 41.32 & 52.26 \\
    $\text{ResNet-Att}$ \cite{li2021lungattn} & - & 71.44 & 36.36 & 53.90 \\
    bi-ResNet-Att \cite{xu2021arsc} & - & 67.13 & {\bf 46.38} & 56.76 \\
    $\text{CNN6}$ \cite{moummad2022supervised} & - & 76.72 & 31.12 & 53.92 \\
    \midrule
    $\text{EfficientNet}^\dagger$ & IN & $\text{\bf 80.07}_{\pm 4.74}$ & $\text{33.05}_{\pm 4.14}$ & $\text{56.56}_{\pm 0.79}$ \\
    $\text{ResNet18}^\dagger$ & IN & $\text{76.63}_{\pm 1.87}$ & $\text{31.06}_{\pm 2.45}$ & $\text{53.84}_{\pm 1.26}$ \\
    ResNeSt \cite{wang2022domain} & IN & 67.80 & 37.10 & 52.20 \\
    $\text{ResNet34}$ \cite{gairola2021respirenet} & IN & 71.40 & 39.00 & 55.20 \\ 
    ResNet50 \cite{nguyen2022lung} & IN & 76.33 & 37.37 & 56.85 \\ 
    $\text{CNN6}$ \cite{moummad2022supervised} & AS & 70.09 & 40.39 & 55.24 \\
    \midrule
    $\text{AST}^\dagger$ & - & $\text{72.61}_{\pm 14.90}$ & $\text{30.58}_{\pm 7.97}$ & $\text{49.60}_{\pm 1.75}$ \\
    $\text{AST}^\dagger$ & IN & $\text{\underline{78.69}}_{\pm 0.60}$ & $\text{38.78}_{\pm 0.65}$ & $\text{\underline{58.73}}_{\pm 0.22}$ \\
    \rowcolor[gray]{0.85}
    $\text{AST}^\dagger$ & \!IN\,+\,AS\! & $\text{77.14}_{\pm 3.35}$ & $\text{\underline{41.97}}_{\pm 2.21}$ & $\text{\bf 59.55}_{\pm 0.88}$ \\
    \bottomrule
    \end{tabular}
    }
    \vspace{-6.5pt}
\end{table}

%% file: tex/04_2_Patch-Mix_Contrastive_Learning.tex
\subsection{Patch-level CutMix on AST Architecture}
\vspace{-2pt}

We have also explored the data augmentation techniques to regularize models with augmented samples.
The Mixup \cite{zhang2017mixup} and Cutmix \cite{yun2019cutmix} methods are popular plug-and-play techniques that can be applied to any modality. Since Transformer-based models split inputs into a sequence of patches (or tokens), recent works have utilized mixing techniques at the patch-level \cite{liu2022tokenmix, choi2022tokenmixup}. To build on this concept, we propose an augmentation technique called Patch-Mix that simply mixes patch embeddings with those from randomly chosen instances, as illustrated in Figure\,\ref{fig:overview}. Our approach does not rely on complex operations such as neural activation maps from a pretrained teacher model \cite{liu2022tokenmix} or guided attention \cite{choi2022tokenmixup}.

Initially, we adopted the same loss form as Mixup or CutMix, generating the label of cross-entropy loss based on the mixing ratio.
Here, we experimented with two augmentation types of Patch-Mix: random selection of masks (``Patch" in Table\,\ref{tab:mixcl}) and masking among the patch groups belonging to the same time domain to mimic Cutmix in raw audio signals (``T-Patch"). 
However, both augmentations did not show the improvement compared to simple fine-tuning results. We hypothesize that the reason is the label hierarchy in respiratory sounds. Linear combinations of ground truth labels may be counter-intuitive in that patients should be diagnosed to \textit{abnormal} with even small amount of crackle or wheeze sounds. Therefore, the conventional mixed loss function using $\lambda$-interpolated labels may not be appropriate for respiratory sound datasets.
\vspace{-3pt}

\input{tables/mixcl}

\begin{figure}[!t]
    \centering
    \includegraphics[width=\linewidth]{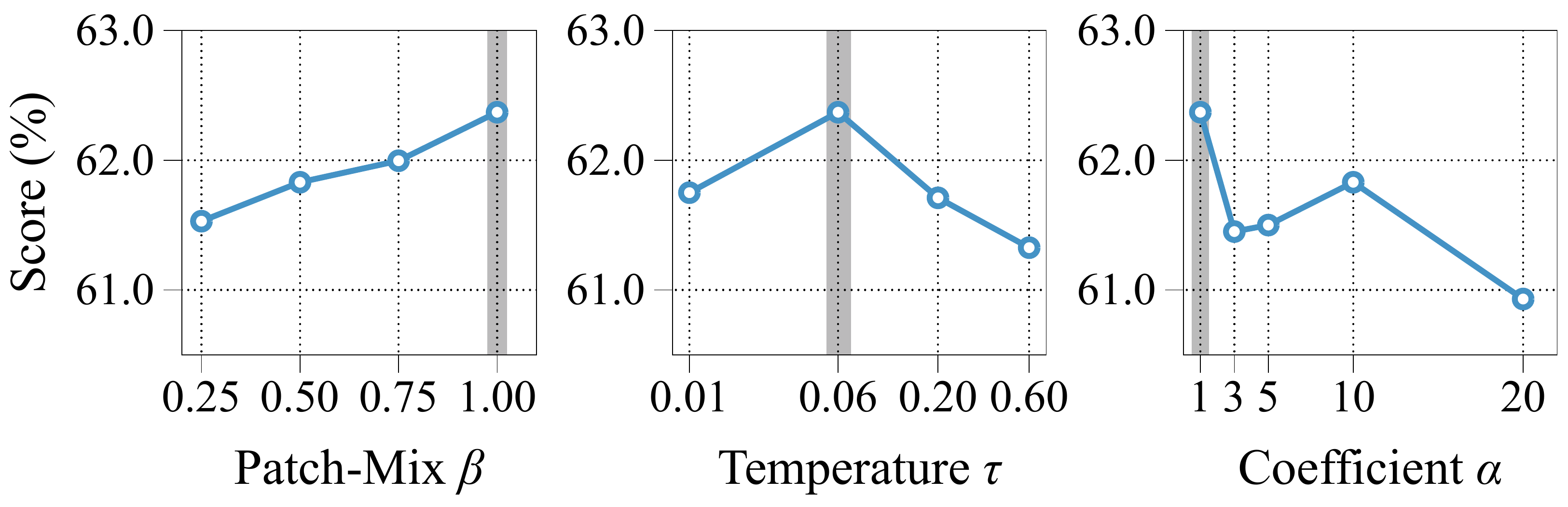}
    \vspace{-15pt}
    \caption{Ablation study on hyperparameters of Patch-Mix contrastive loss. We searched hyperparameters in a default setting: Patch-Mix, ALL negative pair, and $\footnotesize \texttt{stop}(z)$. We highlight the final chosen hyperparameters by gray color.}
    \vspace{-7pt}
    \label{fig:ablation_study}
\end{figure}

\input{tables/main_results.tex}

\subsection{Patch-Mix Contrastive Learning}
\vspace{-2pt}

In order to obtain a suitable loss function, we propose a Patch-Mix contrastive loss that enables the effective learning by distinguishing the mixed information within the latent space:
\vspace{-2pt}
\begin{align}\label{eq:patch_mix_contrastive_loss}
\mathcal{L}_{\text{CL}} \! = -\frac{1}{|I|} \sum_{i \in I} \Big[ \Big( \lambda \! \cdot \! (h(\Tilde{z}_i)^\top z_i &/ \tau) + (1-\lambda) \! \cdot \! (h(\Tilde{z}_i)^\top z_m  / \tau) \Big) \nonumber \\[-5pt]
-\log &\sum_{j\in I} \exp(h(\Tilde{z}_i)^\top z_j  / \tau) \Big]
\vspace{-2pt}
\end{align}
where $z_i$ is the encoder output of patch embedding $E_i$, and $\Tilde{z}_i$ is from mixed embedding with a ratio of $\lambda$ between $E_i$ and $E_m$. $h$ is the projector that consists of two MLP layers with ReLU and BN layers. Note that all representation vectors, such as $h(\Tilde{z}_i)$ or $z_i$, are normalized before the dot product. $I$ is the index set, $\tau$ is the temperature to control the sharpness of cosine similarity, and $\lambda$ is drawn from Beta distribution with parameter $\beta$. 
Here, we formulate a final loss function as follows: $\mathcal{L}_{\text{CE}}$\,+\,$\alpha \mathcal{L}_{\text{CL}}$.
Refer to Figure\,\ref{fig:ablation_study} for the choice of hyperparameters.

Eq.\,\ref{eq:patch_mix_contrastive_loss} is a type of categorical cross-entropy loss, where the numerator (first and second terms) contains the positive pairs while the denominator (third term) consists of both positive and negative pairs.
Our contrastive loss is clearly different to the conventional approach as it considers representations from mixed sources as positive pairs, rather than representations from multi-augmented images \cite{chen2020simple, he2020momentum}. 
Through the mixing ratio $\lambda$, the model learns the degree to which mixed representations should contain the inherent information of the original samples. 

As summarized in Table\,\ref{tab:mixcl}, our experimental findings demonstrate that contrasting Patch-Mixed representations among all batch samples, while applying stop-gradient to the target representations (\textit{e.g.}, $z_i$ or $z_m$), yielded the most promising outcomes. In practice, our proposed Patch-Mix Contrastive Learning significantly improves the Score on the ICBHI dataset by 2.82\% compared to the simple fine-tuning of the AST. 
Given that the conventional mixed loss function did not improve the performance, it proves that our mixed contrastive learning in the latent space is highly effective.

%% file: tables/mixcl.tex
\begin{table}[!t]
    \centering
    \caption{Patch-Mix contrastive learning performance based on three factors: mix augmentations, negative pair types, and target representations. $y$ and $\footnotesize \texttt{stop}$ indicate the label and stop-gradient operation, respectively. \textbf{Best} and {\underline{second best}} results. We highlight the final implementation choices.
    }\label{tab:mixcl}
    \vspace{-2pt}
    \renewcommand{\arraystretch}{1.2}
    \addtolength{\tabcolsep}{-3pt}
    \resizebox{\linewidth}{!}{
    \begin{tabular}{lllllll}
    \toprule
    method & mix aug. & neg. & target & $S_p$\,(\%) & $S_e$\,(\%) & \textbf{Score}\,(\%) \\
    \hline \midrule
    CE & - & - & - & $\text{77.14}_{\pm 3.35}$ & $\text{41.97}_{\pm 2.21}$ & $\text{59.55}_{\pm 0.88}$\\
    CE & Patch & - & - & $\text{76.87}_{\pm 5.05}$ & $\text{{42.04}}_{\pm 3.98}$ & $\text{59.46}_{\pm 0.78}$ \\
    CE & T-Patch & - & - & $\text{79.04}_{\pm 4.88}$ & $\text{40.55}_{\pm 5.14}$ & $\text{59.79}_{\pm 1.20}$ \\
    \midrule
    \rowcolor[gray]{0.85}
    MixCL & Patch & ALL & \texttt{stop}$(z)$ & $\text{\underline{81.66}}_{\pm 3.83}$ & $\text{{43.07}}_{\pm 2.80}$ & $\text{\bf 62.37}_{\pm 0.61}$ \\
    MixCL & T-Patch & ALL & \texttt{stop}$(z)$ & $\text{78.22}_{\pm 7.96}$ & $\text{\bf 44.79}_{\pm 6.17}$ & $\text{61.51}_{\pm 1.15}$ \\
    MixCL & Patch & Diff.\,$y$ & \texttt{stop}$(z)$ & $\text{80.24}_{\pm 4.60}$ & $\text{{40.93}}_{\pm 4.44}$ & $\text{60.59}_{\pm 0.50}$ \\
    MixCL & Patch & ALL & \texttt{stop}$(h(z))$ & $\text{78.07}_{\pm 2.41}$ & $\text{{\underline{43.54}}}_{\pm 1.75}$ & $\text{60.81}_{\pm 0.41}$ \\
    MixCL & Patch & ALL & $z$ & $\text{\bf 82.17}_{\pm 5.52}$ & $\text{{40.87}}_{\pm 4.02}$ & $\text{\underline{61.52}}_{\pm 1.37}$ \\
    MixCL & Patch & ALL & $h(z)$ & $\text{79.89}_{\pm 4.08}$ & $\text{42.06}_{\pm 4.75}$ & $\text{60.97}_{\pm 0.55}$ \\
    \bottomrule
    \end{tabular}
    }
\end{table}

%% file: tables/main_results.tex
\begin{table*}[!t]
    \centering
    \vspace{-4pt}
    \caption{Overall comparison on ICBHI lung sound classification task. We compared to previous works that use the official split of the ICBHI dataset (\textit{i.e.}, 60-40\% split for the train and test set). $*$ denotes the previous state-of-the-art Score. \textbf{Best} and {\underline{second best}} results.}
    \vspace{-5pt}
    \label{tab:main_results}
    \renewcommand{\arraystretch}{1}
    \addtolength{\tabcolsep}{3pt}
    \resizebox{\linewidth}{!}{
    \begin{tabular}{p{2pt}lllllll}
    \toprule
    & method & architecture & pretrain & venue & $S_p$\,(\%) & $S_e$\,(\%) & \textbf{Score}\,(\%) \\
    \hline \midrule
    \multirow{18.5}{*}{\rotatebox[origin=c]{90}{\textbf{4-class eval.}}} & LungBRN \cite{ma2019lungbrn} & bi-ResNet & - & \textit{BioCAS`19} & 69.20 & 31.12 & 50.16 \\
    & SE+SA \cite{yang2020adventitious} & ResNet18 & - & \textit{INTERSPEECH`20} & {81.25} & 17.84 & 49.55 \\
    & LungRN+NL \cite{ma2020lungrn+} & ResNet-NL & - & \textit{INTERSPEECH`20} & 63.20 & 41.32 & 52.26 \\
    & CNN-MoE \cite{pham2021cnn} & C-DNN & - & \textit{JBHI`21} & 72.40 & 21.50 & 47.00 \\
    & LungAttn \cite{li2021lungattn} & ResNet-Att & - & \textit{IPEM`21} & 71.44 & 36.36 & 53.90 \\
    & ARSC-Net \cite{xu2021arsc} & bi-ResNet-Att & - & \textit{BIBM`21} & 67.13 & \textbf{46.38} & 56.76 \\
    & RespireNet \cite{gairola2021respirenet} & ResNet34 & IN & \textit{EMBC`21} & 71.40 & 39.00 & 55.20 \\ 
    & RespireNet \cite{gairola2021respirenet} (CBA+BRC+FT) & ResNet34 & IN & \textit{EMBC`21} & 72.30 & 40.10  & 56.20 \\
    & Ren \textit{et al.} \cite{ren2022prototype} & CNN8-Pt & - & \textit{ICASSP`22} & 72.96 & 27.78 & 50.37 \\
    & Chang \textit{et al.} \cite{chang22h_interspeech} & CNN8-dilated & - & \textit{INTERSPEECH`22} & 69.92 & 35.85 & 52.89 \\
    & Chang \textit{et al.} \cite{chang22h_interspeech} & ResNet-dilated & - & \textit{INTERSPEECH`22} & 50.22 & 51.83 & 51.02 \\
    & Wang \textit{et al.} \cite{wang2022domain} (Splice) & ResNeSt & IN & \textit{ICASSP`22} & 70.40 & 40.20 & 55.30 \\
    & Late-Fusion \cite{9871440} & Inc-03\,+\,VGG14 & IN & \textit{EMBC`22} & \textbf{85.60} & 30.00 & 57.30 \\
    & Nguyen \textit{et al.} \cite{nguyen2022lung}\,(StochNorm) & ResNet50 & IN & \textit{TBME`22} & 78.86 & 36.40 & 57.63 \\
    & Nguyen \textit{et al.} \cite{nguyen2022lung}\,(CoTuning) & ResNet50 & IN & \textit{TBME`22} & 79.34 & 37.24 & $\text{58.29}^\textbf{*}$ \\
    & Moummad \textit{et al.} \cite{moummad2022supervised} & CNN6 & AS & \textit{arXiv`22} & 70.09 & 40.39 & 55.24 \\
    & Moummad \textit{et al.} \cite{moummad2022supervised}\,(SCL) & CNN6 & AS & \textit{arXiv`22} & 75.95 & 39.15 & 57.55 \\
    \cmidrule{2-8}
    & \textbf{Fine-tuning [ours]} & AST & IN\,+\,AS & \textit{INTERSPEECH`23} & $\text{77.14}_{\pm 3.35}$ & $\text{41.97}_{\pm 2.21}$ & $\text{\underline{59.55}}_{\pm 0.88}$ \\
    & \textbf{Patch-Mix CL [ours]} & AST & IN\,+\,AS & \textit{INTERSPEECH`23} & $\text{\underline{81.66}}_{\pm 3.83}$ & $\text{\underline{43.07}}_{\pm 2.80}$ & $\text{\bf 62.37}_{\pm 0.61}$ \\
    \midrule
    \multirow{5.5}{*}{\rotatebox[origin=c]{90}{\textbf{2-class eval.}}} & 
    CNN-MoE \cite{pham2021cnn} & C-DNN & - & \textit{JBHI`21} & 72.40 & 37.50 & 54.10 \\
    & Nguyen \textit{et al.} \cite{nguyen2022lung}\,(StochNorm) & ResNet50 & IN & \textit{TBME`22} & 78.86 & 49.79 & 64.32 \\
    & Nguyen \textit{et al.} \cite{nguyen2022lung}\,(CoTuning) & ResNet50 & IN & \textit{TBME`22} & \underline{79.34} & 50.14 & $\text{64.74}^\textbf{*}$ \\
    \cmidrule{2-8}
    & \textbf{Fine-tuning [ours]} & AST & IN\,+\,AS & \textit{INTERSPEECH`23} & $\text{77.14}_{\pm 3.35}$ & $\text{\bf 56.40}_{\pm 1.73}$ & $\text{\underline{66.77}}_{\pm 1.15}$ \\
    & \textbf{Patch-Mix CL [ours]} & AST & IN\,+\,AS & \textit{INTERSPEECH`23} & $\text{\bf 81.66}_{\pm 3.83}$ & $\text{\underline{55.77}}_{\pm 2.83}$ & $\text{\bf 68.71}_{\pm 0.59}$ \\
    \bottomrule
    \end{tabular}}
    \vspace{-6pt}
\end{table*}

%% file: tex/06_Results.tex
\vspace{-2pt}
\section{ICBHI Dataset Results}
\vspace{-1pt}



Table\,\ref{tab:main_results} presents the overall respiratory sound classification performances of the ICBHI dataset, including our proposed method. 
Based on the overall comparisons, pretrained models on large-scale datasets showed relatively high performance regardless of architectures or methods.
Interestingly, the straightforward fine-tuning of the AST model (``Fine-tuning") achieves a 59.55\% Score, which outperforms the current state-of-the-art model \cite{nguyen2022lung} by 1.26\%. 
In addition, our Patch-Mix Contrastive Learning (``Patch-Mix CL") achieved the state-of-the-art Score in both 4-class and 2-class evaluations, with the values of 62.37\% and 68.71\%, respectively.
We would like to highlight that our methodology demonstrated a significant increase of approximately 3\% compared to simple fine-tuning of the AST, while the methodologies proposed in previous literature have shown only a 1-2\% increase within their architectures. 

%% file: tex/07_Conclusion.tex
\vspace{-3pt}
\section{Conclusion}
\vspace{-1pt}

In this work, we demonstrate the effectiveness of the AST model, a purely attention-based model pretrained on visual and audio datasets, in improving the classification Score of respiratory sounds. 
Additionally, we propose a Patch-Mix augmentation that randomly replaces patches with those from different samples.
However, the conventional mixed loss function did not result in performance gain, likely due to the label hierarchy in lung sounds. Therefore, we introduce a Patch-Mix Contrastive Learning that distinguishes the mixed representations in the latent space.
Our method achieved a new record 62.37\% Score on the ICBHI dataset, outperforming all existing methods.

